\documentclass[amsmath,nofootinbib,amssymb]{revtex4}
\usepackage{rotating}
\usepackage{float}
\usepackage{verbatim}
\usepackage{epsfig}
\usepackage{graphicx}
\usepackage{dcolumn}
\usepackage{bm}
\usepackage{color}
\usepackage{tabularx}
\usepackage{cleveref}
\usepackage{epstopdf}

\newcommand{\Od}{{\cal O}}

\newcommand{\zerof}[1]{\left\lfloor #1 \right\rfloor}

\def\thebiblio#1{
\begin{center}\bf \large References
\end{center}
\list
{[\arabic{enumi}]}{\settowidth\labelwidth{#1.}\leftmargin\labelwidth
 \advance\leftmargin\labelsep
 \usecounter{enumi}}
 \def\newblock{\hskip .11em plus .33em minus -.07em}
 \sloppy
 \sfcode`\.=1000\relax}

\begin{document}
\preprint{}
\title{%
Perturbations of ultralight vector field dark matter
}

\author{J. A. R. Cembranos, A.\,L.\,Maroto and S. J. N\'u\~nez Jare\~no}
\address{Departamento de  F\'{\i}sica Te\'orica I, Universidad Complutense de Madrid, E-28040 Madrid, Spain}

\date{\today}

\begin{abstract}
We study the dynamics of cosmological perturbations in models of dark matter based on ultralight coherent vector fields. Very much as for scalar field dark matter, we find two different 
regimes in the evolution:  for modes with $k^2\ll {\cal H}ma$, we have  a particle-like behaviour
indistinguishable from cold dark matter, whereas for modes with $k^2\gg {\cal H}ma$, 
we get a wave-like behaviour in which the  sound speed is non-vanishing and of order 
$c_s^2\simeq k^2/m^2a^2$. This implies that, also in these models,  
structure formation could be suppressed on small scales. 
However, unlike the scalar case, 
the fact that the background evolution contains a non-vanishing homogeneous vector field implies that,  in general, the evolution of the three kinds of perturbations (scalar, vector and tensor) can no longer be  decoupled at the linear level. More specifically,  in the particle regime, the three
types of perturbations are actually decoupled, whereas in the wave regime,  the three vector field perturbations generate one scalar-tensor and two vector-tensor perturbations
in the metric. Also in the wave regime, 
we find that a non-vanishing anisotropic stress is present in the perturbed energy-momentum tensor
giving rise to a gravitational slip of order $(\Phi-\Psi)/\Phi\sim c_s^2$. Moreover in this regime 
the amplitude of the  tensor to scalar ratio of the scalar-tensor modes is also $h/\Phi\sim c_s^2$. This implies that small-scale density perturbations are necessarily  associated to the presence of gravity waves in this model. We compare their spectrum with the sensitivity of present and future gravity waves detectors.

\end{abstract}

\maketitle

\section{Introduction}

Despite the success of the collisionless cold dark matter (CDM) scenario  in the description of the process of structure formation \cite{Tegmark}, still important 
difficulties are present regarding the predictions of simulations on sub-galactic scales. Indeed, dark matter (DM) only N-body simulations predict cuspy profiles for the DM halo densities whereas observations of DM dominated 
objects, such as dwarf spheroidal galaxies, suggest more cored distributions (cusp-core problem) \cite{cuspcore,cuspcore2,cuspcore3}.
Also, this kind of simulations  predicts more satellite galaxies for Milky Way type objects
than actually observed (missing satellite problem) \cite{missing,missing2}. Finally the central densities of the
most massive simulated subhalos are much higher than
those observed in the most luminous satellite galaxies (too big to fail problem) \cite{toobig}.

Possible solutions to such problems have been suggested in recent years.  In particular the inclusion  of different baryonic physics effects
in the simulations, 
such as feedback from supernova explosions and stellar wind or cosmic ray heating have been proposed among others \cite{cuspcore,baryonic2}.
However, there are other proposals which are based on  the modification of the CDM scenario itself. Thus the interest  
in  warm DM \cite{WDM}, self-interacting \cite{self} or decaying \cite{decaying,decaying2} DM models 
which typically generate a small scale cutoff in the matter power spectrum
has grown in recent years.

Another proposal along these lines is the so called wave DM 
model \cite{Sin,Sin2,Sin3,Sin4,Sin5,Sin6,Sin7,Sin8,Sin9,Sin10,Sin11,Sin12,Sin13,Sin14,Sin15,Sin16,Peebles,Peebles2,Peebles3,Peebles4,Peebles5,Matos-2000,Matos-2000b,Matos-2000c,Matos-2000d,Matos-2000e} which has been also known as 
fuzzy DM, i.e.  scalar field DM 
made of ultralight bosons with negligible selfinteractions. The most popular candidates being the axion-like particles (ALPs) with very small masses typically arising in string theory \cite{string,stringb}. In these wave DM scenarios, it is the uncertainty principle what prevents
the formation of structures on small scales. Indeed, if DM is made of very light particles 
with masses $m\ll 1$ eV, the corresponding number density is so high that the interparticle 
separation becomes smaller than the Compton wavelength so that a field description of 
DM would be possible (Bose-Einstein condensate). 
As a matter of fact, at the background level, i.e. without perturbations,  for massive scalars this field description can be seen as that of a coherently oscillating classical field whose average energy density precisely scales as CDM \cite{Turner}.  
Moreover,  the effect on perturbations of the very light fields can also be understood easily if 
we take into account that
for  masses below $10^{-22}$ eV, 
the de Broglie wavelength of a slowly moving DM particle is of astrophysical size. 
More concretely the comoving de Broglie wavelength is $\lambda_{deBroglie}=({\cal H}ma)^{-1/2}$ \cite{Ferreira}. 
 This means that
since it is not possible to localize the DM particle on scales smaller than $\lambda_{deBroglie}$, 
 structure formation is suppressed on those small scales \cite{nature,natureb,naturec,natured}. 
 Thus, in this kind of models, we have two different regimes for perturbations. On scales larger than $\lambda_{deBroglie}$, the usual particle-like behaviour is a good description and the standard CDM behaviour is recovered, whereas on smaller scales we have a wave-like behaviour which suppresses structure formation.

The wave DM scenario has been considered so far for scalar fields. Thus  
the general analysis of the behaviour of these scalar field models at the background level was developed in \cite{Turner} and the study of its perturbations can be found in \cite{Ratra-1991,Ratra-1991b,Ratra-1991c,Ratra-1991d,Hwang} for massive scalars and \cite{Cembranos-2016} for a generic power-law potential. However, 
 in principle, the scenario could be also implemented for any bosonic field. The main problem which arises in
the case of vectors or higher spin fields is that coherent homogeneous fields 
typically break isotropy. However it has been recently shown (see \cite{Dimopoulos,Isotropy} for abelian vector fields, \cite{IsotropyYM} for non-abelian  theories and \cite{GIsTh} for arbitrary spin) that for rapidly oscillating coherent fields, even though the field evolution is generically anisotropic, the average energy-momentum tensor is not. In particular, for massive fields  it is straightforward to 
show that the average energy density scales as $a^{-3}$.  This opens the possibility 
of extending the wave DM scenario to higher spin fields.

In this work, we will consider the case of massive abelian vector fields.
The interest in homogeneous vector fields as  cosmological fluids has been growing in the last years, see \cite{Galtsov,Galtsovb,Armendariz,Harko,Beltran,Beltranb} for dark energy examples and \cite{vectorinflation,vectorinflationb} for inflation models based on vector fields. 
The possibility that a condensate of very light vector particles could 
play the role of DM was explored  in \cite{Nelson} and a wide phenomenological study of this model was made in \cite{wispy}. Such a condensate could be produced during inflation and its small mass could be generated by the Stuckelberg mechanism. A small kinetic mixing with the 
photon could make this dark photon detectable \cite{darkp1,darkp2,darkp3}.  

Here, in particular, we will concentrate in the dynamics
of cosmological perturbations in such vector DM models. The main difficulty compared to the scalar case is the presence of 
a non-vanishing vector field already at the background level. This implies that the usual decoupling 
of the evolution of scalar, vector and tensor perturbations at the linear level no longer holds in this case.
However, this fact provides a potential way of discriminating vector models from scalar ones. 
In particular, we will show  that although in the particle regime with $k^2\ll {\cal H}ma$
the model is indistinguishable from CDM, in the wave regime $k^2\gg {\cal H}ma$ the scalar and vector modes are coupled to the tensors. This implies that unlike scalar field 
models, density perturbations generate  
a specific gravity wave spectrum together with a non-vanishing anisotropic stress.

The work is organized as follows. Firstly, in Section II, we will review the time averaging procedure in cosmology. Then, in Section III, we will consider the anisotropy problem 
for homogeneous vector fields. In Section IV we obtain the basic equations for 
perturbations of massive vectors and in Section V we write them for
scalar and vector modes. Sections VI and VII are devoted to the results for  scalar and vector perturbations where the adiabatic solutions of the perturbations equations 
are obtained in the different regimes. In Section VIII we concentrate on the 
generation of gravity waves and Section IX in the possibility of detection. Finally Section X includes the main conclusions of the work.

\section{Time averaging in cosmology}
Many of the results we will obtain in this work are based on the assumption that in the presence 
of rapidly oscillating fields, it is possible to time average the energy-momentum tensor so
that the resulting solutions of Einstein equation are a good approximation to the exact ones.
In order to determine when this is the case, we will consider a simple example, which will 
help us to understand the key aspects of this procedure.

Let us consider a homogeneous scalar field oscillating in  a power-law potential 
\begin{eqnarray}
V(\phi)=\frac{\lambda}{n} \phi^n \; ,
\end{eqnarray}
with $n$ and even integer. 
In a flat FLRW metric in proper time,
\begin{eqnarray}
ds^2 = dt^2- a^2(t) d\vec{x}^2\;, 
\end{eqnarray}
the equation of motion  can be written as
\begin{eqnarray}
\ddot{\phi} + 3 \frac{\dot{a}}{a} \dot{\phi} + \lambda \phi^{n-1}=0\;,\label{eom2}
\end{eqnarray}
where the dot represents the $t$ derivative. Making the change $\phi=\tilde{\phi}\; a^{-\frac{6}{n+2}}$ and $dt = a^{\frac{3(n-2)}{n+2}} d\tilde{\eta}$, (\ref{eom2}) reads
\begin{eqnarray}
\tilde{\phi}'' + \lambda \tilde\phi^{n-1}+6\left(\frac{n-4}{(n+2)^2}\left(\frac{a'}{a}\right)^2 - \frac{a''}{(n+2)a}\right)\tilde{\phi}=0\;, \label{eqtilde}
\end{eqnarray}
where $'$ is the derivative with respect to the new time variable $\tilde\eta$. 
Let us assume that the frequency of the oscillations $\omega$ is large
compared to the rate of expansion of the universe i.e. $\omega\gg {\tilde{\cal H}}$
with ${\tilde{\cal H}}=a'/a$. Thus we can define the small parameter $\epsilon\equiv {\tilde{\cal H}}/\omega$. Accordingly, the terms proportional to $\tilde \phi$ in (\ref{eqtilde}) will be suppressed by $\mathcal{O}(\epsilon^2)$ compared to the other ones and we can write
\begin{eqnarray}
\tilde{\phi}'' + \lambda \tilde\phi^{n-1}+ \mathcal{O}(\epsilon^2)=0\;.
\end{eqnarray}
Thus, the solution can be written in terms of the field $\phi$ as
\begin{eqnarray}
\phi(\tilde\eta) = F(\tilde\eta) P(\tilde\eta) + \mathcal{O}(\epsilon^2) \label{phi}\;,
\end{eqnarray}
where $F=a^{-\frac{6}{n+2}}$ is a slowly evolving fuction of $\tilde \eta$
with $F'/F\sim \tilde{\cal H}$ and $P$ is a periodic fast oscillating function with period $2\pi/\omega$, i.e. $P'/P\sim \omega$.

Let us now try to obtain the scale factor $a(\tilde\eta)$ from Einstein equations. The system formed by the Friedmann and conservation equations read
\begin{eqnarray}
H^2 &\equiv& \left(\frac{\dot{a}}{a}\right)^2= \frac{8 \pi G}{3} \rho\;,\\
\dot{\rho} &=& - 3 H \dot{\phi}^2\;,
\end{eqnarray}
from which we can obtain,
\begin{eqnarray}
\dot{H} = - 4 \pi G \dot{\phi}^2 ,
\end{eqnarray}
so that integrating twice in time we get:
\begin{eqnarray}
 a = a_0 \exp\left( - 4 \pi G \int^t_{t_0}dt_1\int^{t_1}_{t_0}  dt_2 \dot{\phi}^2(t_2) \right)=
a_0 \exp\left( - 4 \pi G \int^{\tilde\eta}_{\tilde\eta_0}d\tilde\eta_1a^{\frac{3(n-2)}{n+2}}(\tilde\eta_1)\int^{\tilde\eta_1}_{\tilde\eta_0}  d\tilde\eta_2 a^{\frac{-3(n-2)}{n+2}}(\tilde\eta_2) {\phi'}^2(\tilde\eta_2) \right) . \label{aexact}
\end{eqnarray}
The  $\phi'$ terms in the integrand are dominated by the derivatives of the rapidly oscillating function
so that we can approximate 
\begin{eqnarray}
\phi'^2(\tilde\eta)\simeq F^2(\tilde\eta)P'^2(\tilde\eta)+\Od(\epsilon).
\label{phip}
\end{eqnarray}
 Since 
$P'^2(\tilde\eta)$ is also a periodic function we can Fourier expand it as:
\begin{eqnarray}
P'^2(\tilde\eta)=c_0+\sum_{m=1}^\infty c_m \cos(m\omega\tilde\eta).
 \label{Fourier}
\end{eqnarray}
Let us now perform the first time integration
\begin{eqnarray}
\int^{\tilde\eta_1}_{\tilde\eta_0}\tilde F^2(\tilde\eta_2)P'^2(\tilde\eta_2) d\tilde\eta_2 =
c_0\int^{\tilde\eta_1}_{\tilde\eta_0} \tilde F^2(\tilde\eta_2) d\tilde\eta_2 +
\sum_{m=1}^\infty c_m\int^{\tilde\eta_1}_{\tilde\eta_0}\tilde  F^2(\tilde\eta_2)\cos(m\omega\tilde\eta_2) d\tilde\eta_2\nonumber \\
\end{eqnarray}
with $\tilde F^2(\tilde\eta_2)=a^{\frac{-3(n-2)}{n+2}}(\tilde\eta_2) F^2(\tilde\eta_2)=a^{-3}(\tilde\eta_2)$.
Integrating by parts the $m>0$ terms we get,
\begin{eqnarray}
\int^{\tilde\eta_1}_{\tilde\eta_0}\tilde F^2(\tilde\eta_2)P'^2(\tilde\eta_2) d\tilde\eta_2 &=&
c_0\int^{\tilde\eta_1}_{\tilde\eta_0} \tilde F^2(\tilde\eta_2) d\tilde\eta_2 +
\sum_{m=1}^\infty \left[\frac{c_m\tilde  F^2(\tilde\eta_2)}{m\omega}\sin(m\omega\tilde\eta_2)\right]^{\tilde\eta_1}_{\tilde\eta_0} 
\nonumber \\
&+&
\sum_{m=1}^\infty \left[\frac{c_m\partial_{\tilde\eta_2}{\tilde  F^2}(\tilde\eta_2)}{m^2\omega^2}\cos(m\omega\tilde\eta_2)\right]^{\tilde\eta_1}_{\tilde\eta_0} +\dots = I_0+\sum_{m=1}^\infty I_m\label{pInt}.
\end{eqnarray}
Notice that $\tilde F^2(\tilde\eta_1)$ is proportional to the first derivative of $I_0$ which in general is expected to be, \begin{eqnarray}
\frac{\tilde F^2(\tilde\eta_1)}{\int^{\tilde\eta_1}_{\tilde\eta_0} \tilde F^2(\tilde\eta_2) d\tilde\eta_2} \sim \mathcal{O}(\tilde{\mathcal{H}}).
\end{eqnarray}
Thus we see that compared to the $I_0$ term, the amplitude of the oscillating $I_{m>0}$ contributions are generically suppressed by:
\begin{eqnarray}
\frac{I_{m>0}}{I_0} \sim \mathcal{O}\left(\epsilon\right).
\end{eqnarray}
Moreover, the second integration in (\ref{aexact}), reduces in another $\Od(\epsilon)$ factor the oscillatory contributions.

Notice also that the periodic factor of the  $\mathcal{O}(\epsilon)$ correction  term in (\ref{phip}) can be expressed as a total time derivative, $P'(\tilde\eta)P(\tilde\eta) = \partial_{\tilde\eta} P^2(\tilde\eta)$, which does not contribute to the zero mode of the Fourier expansion, $c_0$. Thus, in general,  we can expand the scale factor as
\begin{eqnarray}
a(\tilde\eta)=a_{m=0}(\tilde\eta)+a_{m>0}(\tilde\eta)= a_{m=0}(\tilde\eta)+\Od(\epsilon^2)
\end{eqnarray}
where $a_{m=0}(\tilde\eta)$ is the contribution from the $c_0$ term whereas $a_{m>0}(\tilde\eta)$
are the oscillatory contributions. We can conclude that, up to $\Od(\epsilon^2)$, it is 
a good approximation to neglect the oscillatory terms $c_{m>0}$ in the source 
of Einstein equations provided the solution involves two time integrations. 
Thus we will denote by $\lfloor\;\rfloor$ the operation of extracting the $m=0$ mode of 
the Fourier expansion, i.e.: 
\begin{eqnarray}
\lfloor a(\tilde\eta)\rfloor=a_{m=0}(\tilde\eta).
\end{eqnarray}
Notice that this operation is equivalent to time averaging $\langle \; \rangle$ up to 
$\Od(\epsilon)$ terms. Indeed
\begin{eqnarray}
\langle \dot\phi^2\rangle=\frac{1}{T}\int^{\tilde\eta_+T}_{\tilde\eta_0}\tilde F^2(\tilde\eta_1)P'^2(\tilde\eta_1) d\tilde\eta_1 +\Od(\epsilon)&=&
c_0\int^{\tilde\eta_0+T}_{\tilde\eta_0} \tilde F^2(\tilde\eta_1) d\tilde\eta_1 +
\Od((\omega T)^{-1})+\Od(\epsilon)= \lfloor \dot\phi^2 \rfloor  +
\Od(\epsilon)
\end{eqnarray}
where in the last step we have considered that both uncertainties are of the same order. 
Consequently, in general if we consider the average Einstein equations
\begin{eqnarray}
G_{\mu \nu} = 8 \pi G \left\lfloor T_{\mu \nu} \right\rfloor,
\end{eqnarray}
the corresponding solutions for $g_{\mu\nu}$ would differ from the exact ones in $\Od(\epsilon^2)$ terms.

\section{Massive vector cosmology}
Let us consider a massive abelian vector field in an expanding universe \cite{GIsTh}. The corresponding action  reads
\begin{equation}
\mathcal{S} = \int d^4x \sqrt{g}\left(- \frac{1}{4} F_{\mu \nu} F^{\mu \nu}- \frac{m^2}{2} A_{\mu} A^{\mu}\right)\;,
\end{equation}
with
\begin{eqnarray}
F_{\mu \nu}=\partial_\mu A_\nu-\partial_\nu A_\mu \;.
\end{eqnarray}
The equations of motion are given by:
\begin{eqnarray}
F^{\mu\nu}_{\;\;\;\; ;\nu} +m^2A^\mu= 0\;.
\label{fieldeq}
\end{eqnarray}
We will first consider the dynamics of the homogeneous background fields. For simplicity we will work with linearly polarized fields
\begin{eqnarray}
A_\mu = \left( A_0(\eta), 0, 0, A_z(\eta)\right).
\label{Azeq}
\end{eqnarray}
Assuming that the energy-momentum tensor is dominated by the vector field, the background geometry can be represented through a Bianchi I metric,
\begin{eqnarray}
ds^2 = a^2 d\eta^2 - a^2 e^{-\frac{b}{2} } dx^2 - a^2 e^{-\frac{b}{2}} dy^2  - a^2 e^{ b} dz^2.
\end{eqnarray}
The $\mu=0$ component of the equation of motion reads
\begin{eqnarray}
m^2A_0=0,
\end{eqnarray}
so that the temporal component identically vanish, whereas 
the $\mu=i$ equations imply
\begin{eqnarray}
\ddot{A}_z - \dot{b} \dot{A}_z + m^2 a^2 A_z=0,
\end{eqnarray}
where dot represents derivative respect to the conformal time $\eta$.
On the other hand, from the exact Einstein equations
\begin{eqnarray}
G_{\mu\nu}=8\pi G T_{\mu\nu}
\end{eqnarray}
we get
\begin{eqnarray}
&&\frac{\ddot{a}}{a} + \frac{\dot{a}^2}{a^2} = \frac{8 \pi G}{3 e^b} \left(\frac{\dot{A}_z^2}{2 a^2} + m^2 A_z^2\right),
\\
&&\ddot{b} + 2 \mathcal{H} \dot{b} = - \frac{32 \pi G}{3 e^b} \left(\frac{\dot{A}_z^2}{2 a^2} - m^2 A_z^2\right),
\\
&&\frac{\dot{a}^2}{a^2} - 16 \dot{b}^2 = \frac{8 \pi G}{3 e^b} \left(\frac{\dot{A}_z^2}{2 a^2} +\frac{m^2}{2} A_z^2\right),
\end{eqnarray}

Let us now assume that the field $A_z$ is oscillating rapidly around the minimum 
of the potential, i.e. we will consider that $ma\gg {\cal H}$, with ${\cal H}=\dot a/a$ the 
comoving Hubble parameter and $ma\gg \dot b$, then the equation of motion
(\ref{Azeq}) can be solved in  the WKB approximation as,
\begin{eqnarray}
A_z = A_{z0}\, a^{-\frac{1}{2}} e^{\frac{b}{2}} \cos\left(\int m a\, d\eta\right)+\mathcal{O}(\epsilon^2),
\end{eqnarray}
with $\epsilon= \{{\cal H}/(ma), \dot b/(ma)\}$. 
Introducing this solution in the system and averaging (extracting the zero mode as discussed in the previous section),  the average Einstein equations
\begin{eqnarray}
G_{\mu \nu} = 8 \pi G \left\lfloor T_{\mu \nu} \right\rfloor,
\label{avEinstein}
\end{eqnarray}
read
\begin{eqnarray}
&&\frac{\ddot{a}}{a} + \frac{\dot{a}^2}{a^2} = 2 \pi G \frac{A_{z0}^2 m^2}{a}  +\mathcal{O}(\epsilon),
\\
&&\ddot{b} + 2 \mathcal{H} \dot{b} = 0+\mathcal{O}(\epsilon),
\\
&&\frac{\dot{a}^2}{a^2} - 16 \dot{b}^2 = \frac{4 \pi G}{3} \frac{A_{z0}^2 m^2}{a} +\mathcal{O}(\epsilon). 
\end{eqnarray}
The second equation shows that there is no source for anisotropy in the average equations, 
so that
\begin{eqnarray}
b(\eta) = b_0 + \frac{b_1}{a^{2}}.
\end{eqnarray}
Thus, if the initial conditions are isotropic then $b(\eta)=0$ at all times  and the third equation reads
\begin{eqnarray}
\frac{\dot{a}^2}{a^2}= \frac{4 \pi G}{3} \frac{ A_{z0}^2 m^2}{a} +\mathcal{O}(\epsilon),
\end{eqnarray}
with solution to leading order in $\epsilon$
\begin{eqnarray}
a = a_0 \left(\frac{\eta}{\eta_0}\right)^2.
\end{eqnarray}
Thus, as shown in \cite{GIsTh} the average  geometry generated by a rapidly oscillating massive  abelian vector field is isotropic and evolves as in a matter dominated universe. If the vector field is responsible for all the DM contribution, its amplitude will be given by
\begin{eqnarray}
A_{z0} =\frac{\sqrt{2\,\Omega_c\, \rho_c}}{m},
\end{eqnarray}
with $\Omega_c$ the CDM density parameter and $\rho_c= 3H_0^2/(8\pi G)$ the 
critical density.

\section{Perturbations of massive vectors.}

In the previous section we have shown that despite the anisotropic evolution of the 
background vector field, the average geometry can be described by an isotropic FLRW metric. 
Thus, we will consider the most general form of the perturbations around the Robertson-Walker 
geometry
\begin{eqnarray}
ds^2 &=& a(\eta)^2 \left[ \left( 1 + 2 \Phi(\eta, \vec{x}) \right) d\eta^2 - \left( \left( 1 - 2 \Psi(\eta, \vec{x}) \right) \delta_{i j} + h_{i j}(\eta, \vec{x})\right) dx^i dx^j - 2 Q_i(\eta, \vec{x}) d\eta dx^i \right]\;, \nonumber
\\
\nonumber
\\
A_{\mu} &=& \left ( \delta A_0(\eta,\vec{x}), \; \vec{A}(\eta) + \delta \vec{A}(\eta,\vec{x}) \right)\;,
\end{eqnarray}
where $\vec{Q}$ is a solenoidal vector field and $h_{i j}$ a symmetric traceless transverse tensor.

From (\ref{fieldeq}), Fourier transforming the spatial dependence, the equation of motion for $\delta A_i$ results,
\begin{eqnarray}
\ddot{\delta A}_i &+& i k_i \dot{\delta A_0} - \left(\dot{\Phi} +\dot{\Psi}\right)\dot{A}_i - 2 \Phi \ddot{A}_i  - i \left( \vec{k}\dot{\vec{A}}\right) Q_i - \dot{h}_{i j} \dot{A}_j + \left(m^2a^2+k^2\right) \delta A_i - k_i \left(\vec{k} \vec{\delta A}\right) =0\;,\;\;\;\;\;\;\; \label{EqdeltaAi}
\end{eqnarray}
and $\delta A_0$ satisfies the constraint
\begin{eqnarray}
\delta A_0( \eta, \vec{k} )= \frac{i \vec{k}\dot{\vec{\delta A}} - i \vec{k}\dot{\vec{A}} \left( \Psi + \Phi\right) + m^2 a^2 \vec{A}\vec{Q}}{m^2 a^2 + k^2} \;.\label{deltaA0}
\end{eqnarray}

The first order perturbations of the  energy-momentum tensor can be written  in Fourier space as
\begin{eqnarray}
\delta T^{\mu}_{\;\nu}(\eta, \vec k) &=& \left[-\frac{\Psi}{a^4} \left(\dot{\vec{A}}^2 - m^2 a^2 \vec{A}^2 \right) + \frac{\Phi}{a^4}\dot{\vec{A}}^2 - \frac{\dot{\vec{A}}\dot{\vec{\delta A}}}{a^4} -i \frac{\vec{k}\dot{\vec{A}}}{a^4} \delta A_0 + \frac{m^2}{a^2} \vec{A}\vec{\delta A} + h_{l m} \left(\frac{\dot{A}_l\dot{A}_m}{2 a^4} + \frac{m^2}{2 a^2} A_l A_m \right)\right] \delta^{\mu}_{\nu}
\\
&+& \delta^\mu_0 \delta^0_\nu \left( 2 \left( \Psi - \Phi \right)\frac{\dot{\vec{A}}^2}{a^4} + 2 i \frac{\vec{k}\dot{\vec{A}}}{a^4} \delta A_0 + 2 \frac{\dot{\vec{\delta A}}\dot{\vec{A}}}{a^4} - h_{l m} \frac{\dot{A}_l\dot{A}_m}{a^4} \right) \nonumber
\\
&+& \delta^\mu_i \delta^j_\nu \left( 2 \left( \Psi - \Phi \right)\frac{\dot{A}_i\dot{A}_j}{a^4} + 2 \frac{\dot{\delta A}_{(i}\dot{ A}_{j)}}{a^4} + 2 i \frac{k_{(i} \dot{A}_{j)}}{a^4}\delta A_0 - 2 \Psi m^2 \frac{A_i A_j}{a^2} - 2 m^2 \frac{\delta A_{(i} A_{j)}}{a^2} + h_{i l} \left(\frac{\dot{A}_j\dot{A}_l}{a^4} + \frac{m^2}{a^2} A_j A_l \right)\right)\nonumber
\\
&+& \delta^\mu_0 \delta^i_\nu \left( i \frac{\vec{k}\dot{\vec{A}}}{a^4} \delta A_i - i k_i \frac{\dot{\vec{A}}\vec{\delta A}}{a^4} - m^2 \frac{\vec{Q}\vec{A}}{a^2} A_i + m^2\frac{A_i \delta A_0}{a^2}\right)\nonumber \\
 &+& \delta^\mu_i \delta^0_\nu \left( \frac{\vec{Q} \dot{\vec{A}}}{a^4}\dot{A}_i-\frac{Q^i}{a^4}\dot{\vec{A}}^2 - i \frac{\vec{k}\dot{\vec{A}}}{a^4}\delta A_i + i \frac{\vec{\delta A} \dot{\vec{A}}}{a^4} k_i - m^2\frac{A_i \delta A_0}{a^2}\right)\;, \nonumber
\end{eqnarray}
and the corresponding perturbations of the average Einstein equations  (\ref{avEinstein}) read
\begin{eqnarray}
&& - 3 \mathcal{H} \left( \dot{\Psi} + \mathcal{H} \Phi\right) - k^2 \Psi = 4 \pi G a^2 \left\lfloor\delta T^0_{\;0} \right\rfloor \;, \label{Ec00}
\\
\nonumber
\\
&& \left[ - 2 \ddot{\Psi} - 2 \left( \mathcal{H}^2 + 2 \dot{\mathcal{H}}\right) \Phi - 2 \mathcal{H} \dot{\Phi}- 4 \mathcal{H} \dot{\Psi}  + k^2 \left(\Phi - \Psi\right) \right] \delta^i_j  + k_i k_j \left( \Psi - \Phi\right)\nonumber
\\
&& - \frac{1}{2} \left( \ddot{h}_{i j} + 2 \mathcal{H} \dot{h}_{i j}+ k^2 h_{i j} \right)- i k_{(i} \dot{Q}_{j)} - i 2 \mathcal{H} k_{(i}Q_{j)} = 8 \pi G a^2 \left\lfloor\delta T^i_{\;j} \right\rfloor\;,\label{Ecij}
\\
\nonumber
\\
&& - 2 i k_i \left( \dot{\Psi} + \mathcal{H} \Phi\right) + \; \frac{k^2}{2}\;Q_{i} = 8 \pi G a^2 \left\lfloor\delta T^0_{\;i}\right\rfloor\;.\label{Ec0i}
\end{eqnarray}
Thus, we can define the average energy density and pressure as:
\begin{eqnarray}
\delta \rho( \eta, \vec{k} ) =\left\lfloor\delta T^0_{\;0}( \eta, \vec{k} ) \right\rfloor,\\
\delta p( \eta, \vec{k} )=\frac{1}{3}\left\lfloor\delta T^i_{\;i}( \eta, \vec{k} ) \right\rfloor.
\end{eqnarray}

Unlike the scalar field case, the perturbations of a vector field can source the three kinds of 
perturbations. This means that the standard separation in the evolution can be more involved in this case. In this work we will proceed as follows: we will first consider the dynamics  
of scalar and vector modes neglecting the contributions from gravity waves. We will
then analyze the generation of gravity waves and will find that they are generically 
suppressed compared to the scalar and vector modes, thus proving that our initial assumption was
correct. 

\section{Scalar and vector perturbations:  basic formulae and preliminaries}

From equations (\ref{Ec00}-\ref{Ec0i}) setting $h_{ij}=0$ we can obtain the following set
of equations, which together with the equation of motion (\ref{EqdeltaAi}) will be the starting point of our analysis:
\begin{itemize}

\item From the combination $\delta G^i_i- 3 \hat{k}_i \hat{k}^j  \delta G^i_j$ we get,
\begin{eqnarray}
k^2 \left( \Psi - \Phi \right) &=& \frac{8\pi G}{a^2} \left\lfloor -\Psi \left( \dot{A}^2 - m^2 a^2 A^2 \right) + \Phi \dot{A}^2 - \dot{\vec{A}} \;\dot{\overrightarrow{\delta A}} - i \delta A_0 \vec{k} \dot{\vec{A}} + m^2 a^2 \vec{A}\;\overrightarrow{\delta A} \right. \label{SlipEq}
\\
&+& \left. 3 \left( \Psi \left( \left(\hat{k}\dot{\vec{A}}\right)^2 - m^2 a^2 \left(\hat{k}\vec{A}\right)^2 \right) - \Phi \left(\hat{k}\dot{\vec{A}}\right)^2  + \left( \hat{k} \dot{\vec{A}}\right) \left( \hat{k} \; \dot{\overrightarrow{\delta A}}\right) + i \delta A_0 \vec{k} \dot{\vec{A}} - m^2 a^2 \left( \hat{k} \;\vec{A} \right) \left( \hat{k} \; \overrightarrow{\delta A} \right)\right)\right\rfloor,\nonumber
\end{eqnarray}
with $\hat{k}$ the unitary vector in the wavenumber direction.
\item From $G^0_0$ we obtain
\begin{eqnarray}
- 3 \mathcal{H} \left( \dot{\Psi} + \mathcal{H} \Phi \right) - k^2 \Psi = \frac{4\pi G}{ a^2} \left\lfloor\Psi \left( \dot{A}^2 + m^2 a^2 A^2 \right) - \Phi \dot{A}^2 + \dot{\vec{A}} \;\dot{\overrightarrow{\delta A}} + i \delta A_0 \vec{k} \dot{\vec{A}} + m^2 a^2 \vec{A}\;\overrightarrow{\delta A} \right\rfloor\;. \label{Eq00}
\end{eqnarray}
\item From the longitudinal part of $G^0_i$ we get
\begin{eqnarray}
\dot{\Psi} + \mathcal{H} \Phi &=& \frac{4\pi G}{a^2} \left\lfloor - \left(\hat{k}\; \dot{\vec{A}}\right)\left(\hat{k}\;\overrightarrow{ \delta A}\right) + \dot{\vec{A}}\; \overrightarrow{\delta A} - i \frac{m^2 a^2}{k} \left(\vec{Q}\vec{A}\right) \left(\hat{k} \vec{A}\right) + i \frac{m^2 a^2}{k} \delta A_0 \left(\hat{k} \vec{A}\right) \right\rfloor\;.\label{Eq0iSca}
\end{eqnarray}
\item From the solenoidal part of $G^0_i$ we can write
\begin{eqnarray}
\vec{Q}( \eta, \vec{k} ) &=& - \frac{16\pi G}{k^2 a^2} i \left\lfloor \vec{k}\dot{\vec{A}} \left(\hat{k}\left(\hat{k}\vec{\delta A}\right) -\vec{\delta A}\right) + m^2 a^2 \left(\vec{Q}\vec{A} - \delta A_0\right) \left(\hat{k}\left(\hat{k}\vec{A}\right) - \vec{A}\right)\right\rfloor\;. \label{Eq0iVec}
\end{eqnarray}
\end{itemize}

Before analysing the modes in the different regimes, we would like to 
make the following preliminary considerations:

\begin{enumerate}
\item For simplicity we will consider a linearly polarized background field, which can 
be written without loss of generality as
\begin{eqnarray}
\vec{A}(\eta) = \vec A_B(\eta)  \cos \left(\int m a \,d\eta \right) +\Od(\epsilon^2);
\end{eqnarray}
with $\vec A_B(\eta)$ a slowly varying amplitude. 

We will work  in the matter dominated era, assuming that all 
the DM is generated by the vector field and ignoring for 
simplicity the 
small baryon contribution. Thus, from the Friedmann equation,
\begin{eqnarray}
\mathcal{H}^2 = \frac{8\pi G}{3}a^2\rho
\end{eqnarray}
with $\rho$ the average energy density:
\begin{eqnarray}
\rho=\left\lfloor\frac{\dot{A}^2}{2a^4}+ \frac{m^2 A^2}{2a^2} \right\rfloor
\end{eqnarray}
we get
\begin{eqnarray}
 A_B (\eta) =\sqrt{\frac{3}{ m^2\eta^2\pi G}}\left(
1 + \mathcal{O}(\epsilon)\right)\;.
\end{eqnarray}

\item As we are dealing with vector equations, it is very helpful to adopt the orthonormal basis $\left\{\hat{u}_a,\; \hat{u}_{pk},\; \hat{u}_p \right\} \equiv \left\{ \hat{u}_A,\; (\hat{k} \times \hat{u}_A)/\sin\theta,\; \left( \hat{k} - \cos \,\theta\; \hat{u}_A\right)/ \sin\theta \right\}$, for modes with $\hat{k}\nparallel \hat{u}_A$; where $\hat{u}_A$ is the normalized vector in $\vec{A}$ direction and $\cos \theta \equiv \hat{k} \cdot \hat{u}_A$. On the other hand, in the degenerate case with $\hat{k}\parallel \hat{u}_A$, we can use a new orthonormal basis $\left\{ \hat{u}_a,\; \hat{u}_{pk1},\; \hat{u}_{pk2} \right\}$, where  
$\{\hat{u}_{pk1},\; \hat{u}_{pk2}\}$ span the orthogonal plane to $\hat{u}_a$. It can be seen that the perturbations in those directions are purely vector with the same dynamics as the $\hat{u}_{pk}$, whereas the $\hat{u}_a$ components
generate purely scalar modes also with the same behaviour as in the 
non-parallel case. Finally, no tensor modes are sourced for $\hat{k}\parallel \hat{u}_A$.

\item The $\hat{u}_a$ component of (\ref{Eq0iVec}) gives us an algebraic equation for $Q_a$. Once it is solved, it is straightforward to write the other two components of $\vec Q$ as a function of the scalar perturbations of the metric and the vector field. After that, combining equations (\ref{SlipEq}), (\ref{Eq00}) and (\ref{Eq0iSca}), we reach an algebraic system from which we obtain $\Psi$ and $\Phi$ depending on $A$ and $\delta A$.

\item We have three independent (comoving) scales in the problem, namely, $ma$, ${\cal H}$ and $k$.
The main assumption of this work  is that $ma\gg \left\{ {\cal H}, k\right\}$, so that we can define two small
parameters $\epsilon={\cal H}/(ma)$ and $k/(ma)$.  Depending on the relation between these two ratios, the evolution of the perturbations will behave differently. Thus, as we will show, the case $k/(ma) \sim \epsilon$,  i.e. $k\sim {\cal H}$ will lead to the standard CDM behaviour, whereas
the  $k/(ma)\sim \epsilon^{1/2}$ case, i.e. $k\sim ({\cal H}ma)^{1/2}$ will correspond to the wave DM behaviour as commented above. We will perform an expansion 
in $\epsilon$ and obtain only the leading order term. The sub-leading correction will 
in general receive contributions from the oscillating terms ($m>0$) mentioned in Section II and are beyond the scope of this work.

\item Provided $k \ll m a$, as mentioned before,  we will take for the 
perturbations an adiabatic 
ansatz similar to that of the background:
\begin{eqnarray}
{\delta \vec A}( \eta, \vec{k} ) = \delta \vec A_s( \eta, \vec{k} )\sin\left(
\int m a d\eta\right)+ \delta \vec A_c( \eta, \vec{k} ) \cos \left(
\int m a d\eta\right)+\Od(\epsilon^2)\;,
\end{eqnarray}
with $\delta \vec A_{(s,c)}$ slowly evolving amplitudes. 
On the other hand, in the regime with $k\sim ma$, the perturbed field oscillates with a different frequency and as a result in the averaging procedure all the perturbed quantities vanish, consequently a cut off in perturbations is expected in the 
high wavenumber region.

\item In the very-low wavenumber regime with $k/(ma) \sim \epsilon^{3/2}$ i.e. $k\sim ({\cal H}^3/(ma))^{1/2}$ the leading order equations get contributions from the oscillating terms that cannot be neglected. Thus, our perturbative approach does not allow to explore this region.  Fortunately, for the masses usually considered \cite{wispy} this range is out of the cosmologically observable band.
\end{enumerate}

\begin{figure*}[h]
\includegraphics[width=0.65 \textwidth]{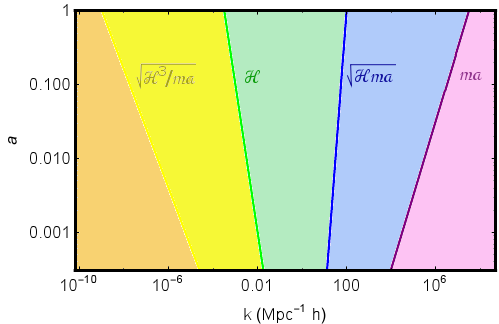}
\caption{Evolution of comoving scales from matter-radiation equality for  $m=10^{-22}$ eV. 
The blue line sets the limit between the particle and wave regimes.
The yellow region corresponds to super-Hubble modes. 
The green one corresponds to sub-Hubble modes. The blue area is the 
wave regime and the pink one is the cutoff region. The orange region on the left  corresponds to the region where the perturbative approach breaks down.}\label{scales}
\end{figure*}

In Fig.\ref{scales} we show the evolution of the different comoving scales involved in the adiabatic expansion, namely,
$({\cal H}^3/(ma))^{1/2}$, ${\cal H}$, $\sqrt{{\cal H}ma}$ and $ma$ as a 
function of $a$ from matter-radiation equality to the present time. We see that modes in 
the wave regime can cross into the particle regime, but this is not 
possible in the opposite way.

\section{Scalar perturbations: results}
As mentioned above, we will concentrate in two different regimes, namely, $k/(ma) \sim \epsilon$ and $k/(ma) \sim \epsilon^{1/2}$. We will present the results of the perturbative analysis to the leading order in the adiabatic expansion, i.e. 
up to relative corrections of order $\epsilon$.

\subsection{Particle regime ($k\sim {\cal H}$, i.e.  $k/(ma) \sim \epsilon$)}
Solving the set of equations (\ref{EqdeltaAi}), (\ref{SlipEq}), (\ref{Eq00}), (\ref{Eq0iSca}), (\ref{Eq0iVec}), we get for the amplitude vectors $\delta \vec A_s(\eta)$ and $\delta \vec A_c(\eta)$ the following solutions in components to leading order.

For the components  orthogonal to the background vector $i = p,\; pk$
 the solution is straightforward,
\begin{eqnarray}
\delta A_{i,(s,c)} ( \eta, \vec{k} ) = a^{-1/2}C_{i, (s,c)} ( \vec{k} )
\end{eqnarray}
with  $C_{i, (s,c)}$ constants. These components do not contribute to the scalar perturbations $\Phi$ and $\Psi$ nor to the density and pressure perturbations. 

The equations for the $i=a$ component  read: 
\begin{eqnarray}
&& \delta A_{a, c} = \frac{\mathcal{H}\left( 48 + k^2 \eta^2 \right)}{24 m a} \delta A_{a, s} + \frac{12 + k^2 \eta^2}{12 m a} \dot{\delta A}_{a, s} \;,
\\
&& \ddot{\delta A}_{a, s} + \mathcal{H} \dot{\delta A}_{a, s} - \frac{3}{2} \mathcal{H}^2 \delta A_{a, s} = 0\;,
\end{eqnarray}
Solving we obtain,
\begin{eqnarray}
\delta A_{a, s}( \eta, \vec{k} ) &=& a \,C_{a 1}( \vec{k} ) + a^{-3/2} C_{a 2}(\vec{k} ) \;,
\\
\Psi( \eta, \vec{k} ) &=& - \sqrt{3\pi G} \,C_{a 1}( \vec{k} )+ \sqrt{\frac{4\pi G}{3}}C_{a 2}( \vec{k} ) \, a^{-5/2} \;,
\\
\Phi( \eta, \vec{k} ) &=&  \Psi( \eta, \vec{k} )\;,
\\
\delta( \eta, \vec{k} ) &=&\frac{\delta \rho( \eta, \vec{k} )}{\rho} = \frac{1}{\sqrt{2}} \left(\left(3 + \frac{k^2\eta^2}{4}\right) C_{a 1}( \vec{k} )+ \left(9 - \frac{k^2\eta^2}{2}\right) a^{-5/2} C_{a 2}(\vec{k} )\right)\;,
\\
\delta p( \eta, \vec{k} ) &=& 0 \;.
\end{eqnarray}
The behaviour is the same as that of standard CDM. Notice that the 
non-decaying mode of the scalar perturbation $\Phi$ is constant independently of the mode and the gravitational slip vanishes since $\Phi=\Psi$. Moreover, the perturbed energy density is controlled by $k^2 \eta^2$, making the density contrast $\delta$ constant for super-Hubble modes and growing as $\delta \sim a$ for sub-Hubble modes as expected.

\subsection{Wave regime ($k\sim ({\cal H}ma)^{1/2}$ i.e.  $k/(ma)\sim \epsilon^{1/2}$)}
This case corresponds to modes whose wavelength is comparable to the 
de Broglie wavelength of  a comoving DM particle. In this regime the 
wave properties of DM could have important effects. 

As in the previous case, we study the evolution of the different components. 
For the $i=p,\; pk$ components, we get to the leading order
\begin{eqnarray}
\delta A_{i,s}( \eta, \vec{k} ) = a^{-1/2} \left(C_{i 2}( \vec{k} ) \sin\left(\frac{k^2}{m a \mathcal{H}}\right)-C_{i 1}( \vec{k} ) \cos\left(\frac{k^2}{m a \mathcal{H}}\right)\right)
\end{eqnarray}
and
\begin{eqnarray}
\delta A_{i,c}( \eta, \vec{k} ) = a^{-1/2} \left(C_{i 1}( \vec{k} ) \sin\left(\frac{k^2}{m a \mathcal{H}}\right)+ C_{i 2}( \vec{k} ) \cos\left(\frac{k^2}{m a \mathcal{H}}\right)\right)
\end{eqnarray}
with $C_{i 1}( \vec{k} )$ and $C_{i 2}( \vec{k} )$  constants. Again these components do not contribute to the scalar perturbations of the metric.

In order to solve for the $\hat{u}_a$ component, we will use equation (\ref{Eq0iSca}) and  average $\langle \vec{A} \cdot$ (\ref{EqdeltaAi}) $\rangle$  obtaining
\begin{eqnarray}
&& \Psi( \eta, \vec{k} ) = -  2\sqrt{3\pi G}\; \frac{m  \mathcal{H}}{k^2}\; \delta A_{a,c}( \eta, \vec{k} )\;, \label{psisolk}
\\
&& \Phi( \eta, \vec{k} )= - 2\sqrt{3\pi G} \; \frac{m  \mathcal{H}}{k^2}\; \delta A_{a,c}( \eta, \vec{k} ) \;,\label{phisolk}
\\
&& \delta A_{a, s} = 2 \frac{m a}{k^2} \left( \dot{\delta A}_{a, c} + \frac{\mathcal{H}}{2} \delta A_{a, c} \right) \;,
\\
&& \ddot{\delta A}_{a, c} + 2 \mathcal{H} \dot{\delta A}_{a, c} + \left( \frac{k^4}{4 m^2 a^2} - \mathcal{H}^2 \right)\delta A_{a, c} = 0\;. \label{ke1}
\end{eqnarray}
By solving (\ref{ke1}) we get,
\begin{eqnarray}
\delta A_{c, a}( \eta, \vec{k} ) = a^{-1/2} && \left[\left(C_{a 2}( \vec{k} ) \left(1 - \frac{3 m^2 a^2 \mathcal{H}^2}{k^4}\right) + 3 C_{a 1}( \vec{k} ) \frac{m a \mathcal{H}}{k^2}\right) \cos \left(\frac{k^2}{m a \mathcal{H}}\right) \right. \label{solk}
\\
&& \left. + \left(C_{a 1}( \vec{k} ) \left(1 - \frac{3 m^2 a^2 \mathcal{H}^2}{k^4}\right) - 3 C_{a 2}( \vec{k} ) \frac{m a \mathcal{H}}{k^2}\right) \sin \left(\frac{k^2}{m a \mathcal{H}}\right)\right]\;. \nonumber
\end{eqnarray}

\begin{figure}
\includegraphics[width = 0.65 \textwidth]{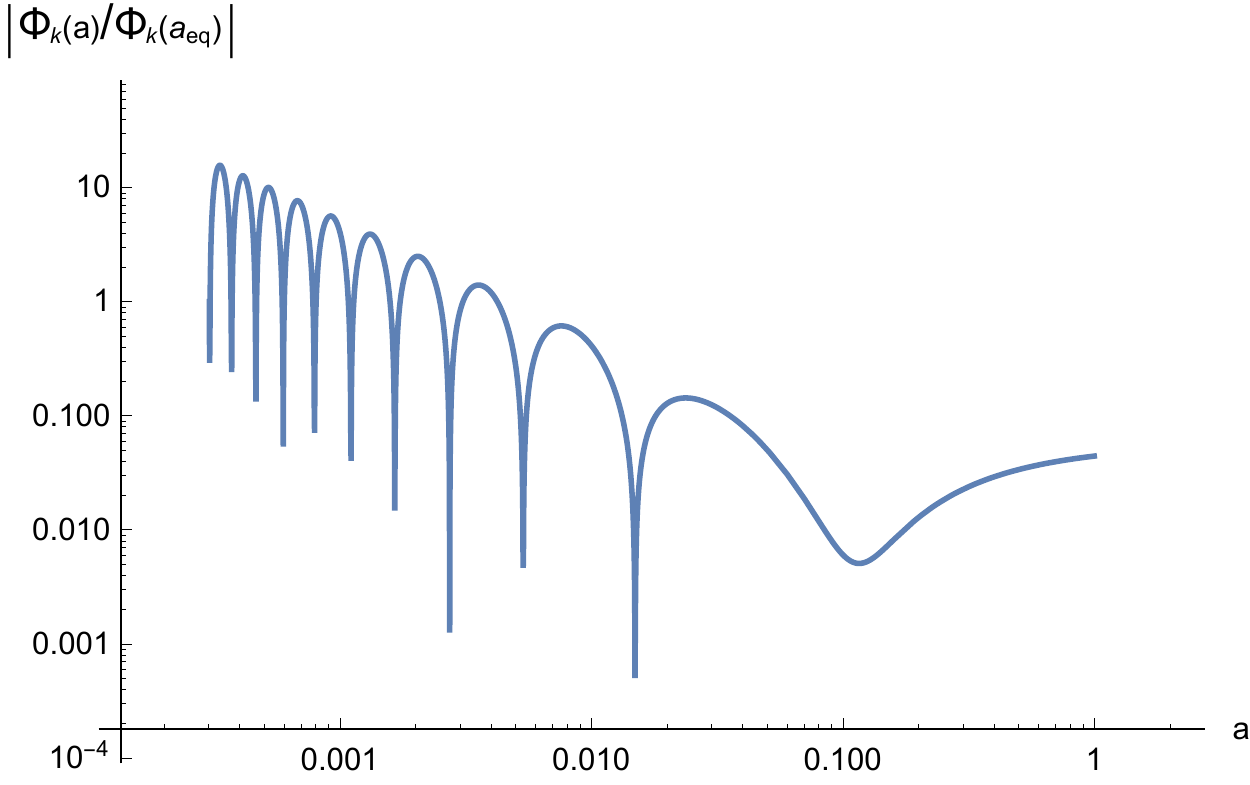}
\caption{Evolution of the $C_{a2}$ mode of the $\Phi$ perturbation with $k=80$ $h$ Mpc$^{-1}$ for a vector mass $m=10^{-22}$ eV normalized to 
the value at matter-radiation equality. We see the decaying oscillating behaviour at early times and the constant asympotic behaviour at late times.}\label{FigScalar}
\end{figure}

The expressions of the metric scalar perturbations are trivially deduced from (\ref{psisolk}), (\ref{phisolk}) and the leading order solution (\ref{solk}). The perturbed energy density and pressure can be written as
\begin{eqnarray}
\delta \rho( \eta, \vec{k} ) &=& \sqrt{\frac{3}{8\pi G}}\frac{m \mathcal{H}}{a^{5/2}} \sqrt{2}  \left[ \left(  \left(1 - \frac{3 m^2 a^2 \mathcal{H}^2}{k^4}\right)C_{a 1}( \vec{k} ) - 3   \frac{m a \mathcal{H}}{k^2} C_{a 2}( \vec{k} )\right) \sin\left(\frac{k^2}{m a \mathcal{H}}\right) \right.
\nonumber \\
&+& \left. \left(  \left(1 - \frac{3 m^2 a^2 \mathcal{H}^2}{k^4}\right)C_{a 2}( \vec{k} ) + 3   \frac{m a \mathcal{H}}{k^2} C_{a 1}( \vec{k} )\right) \cos\left(\frac{k^2}{m a \mathcal{H}}\right)\right]\;,
 \\
 \nonumber
 \\
 \delta p( \eta, \vec{k} ) &=& -\sqrt{\frac{3}{8\pi G}}\frac{\mathcal{H}^2}{a^{7/2}}\frac{k^2}{2 \sqrt{2}m a \mathcal{H}}\cos (2\theta)\left[ \left( \left( 1 - \frac{3m^2 a^2 \mathcal{H}^2}{k^4}\right)C_{a 1}( \vec{k} ) - 3   \frac{m a \mathcal{H}}{k^2} C_{a 2}( \vec{k} ) +\tan(2\theta) \,C_{p 1}( \vec{k} )\right)  \sin\left(\frac{k^2}{m a \mathcal{H}}\right)\right. \nonumber
 \\
&+& \left. \left(\left( 1 - \frac{3m^2 a^2 \mathcal{H}^2}{k^4}\right)C_{a 2}( \vec{k} ) + 3   \frac{m a \mathcal{H}}{k^2} C_{a 1}(  \vec{k} )+\tan(2\theta) \, C_{p 2}( \vec{k} )\right)  \cos\left(\frac{k^2}{m a \mathcal{H}}\right)\right]\;,\label{deltapvec}
\end{eqnarray}
whereas for the scalar metric perturbations we get:
\begin{eqnarray}
\Phi( \eta, \vec{k} ) &=& - 2\sqrt{3\pi G} \; \frac{m  \mathcal{H}}{k^2a^{1/2} } \left[\left(C_{a 2}( \vec{k} ) \left(1 - \frac{3 m^2 a^2 \mathcal{H}^2}{k^4}\right) + 3 C_{a 1}( \vec{k} ) \frac{m a \mathcal{H}}{k^2}\right) \cos \left(\frac{k^2}{m a \mathcal{H}}\right) \right. 
\\
&& \left. + \left(C_{a 1}( \vec{k} ) \left(1 - \frac{3 m^2 a^2 \mathcal{H}^2}{k^4}\right) - 3 C_{a 2}( \vec{k} ) \frac{m a \mathcal{H}}{k^2}\right) \sin \left(\frac{k^2}{m a \mathcal{H}}\right)\right]\;. \label{Phik}\nonumber
\end{eqnarray}
The evolution of the scalar perturbation potential is shown in Fig. \ref{FigScalar}. 

We see that for the $i=p$ component it is not possible to define 
a sound speed.  For  the $i=a$ component  the effective sound speed takes a very simple form,
\begin{eqnarray}
c_{eff}^2 \equiv \frac{\langle \delta p \rangle}{\langle \delta \rho \rangle} = -\frac{k^2}{4 m^2 a^2} \cos(2\theta)\;,
\end{eqnarray}
however, this expression can become negative. As a matter of fact, since
as we will show below, there is a non-vanishing gravitational slip, this is not going to be the characteristic propagation velocity  of scalar perturbations.

Even though to the leading order we get $\Phi=\Psi$, it is possible to derive the sub-leading contribution  from
(\ref{SlipEq})
\begin{eqnarray}
&& \Psi - \Phi = - \sqrt{\frac{8\pi G}{3}}\frac{3 \mathcal{H}}{2 \sqrt{2} m a^2} a^{-1/2} \left[ \left(C_{a 2}( \vec{k} ) \left(1 - 3 \frac{m^2 a^2 \mathcal{H}^2}{k^4}\right) \left(1 + \cos^2 \theta\right) +\frac{1}{2}C_{p 2}( \vec{k} ) \sin(2\theta) + 3 C_{a 1}( \vec{k} ) \frac{m a \mathcal{H}}{k^2} \left(1 + \cos^2 \theta\right)\right)\right.\nonumber
\\
&& \left.\cos\left(\frac{k^2}{m a \mathcal{H}}\right)+\left(C_{a 1}( \vec{k} ) \left(1 - 3 \frac{m^2 a^2 \mathcal{H}^2}{k^4}\right) \left(1 + \cos^2 \theta\right) + \frac{1}{2}C_{p 1}( \vec{k} ) \sin(2\theta) - 3 C_{a 2}( \vec{k} ) \frac{m a \mathcal{H}}{k^2} \left(1 + \cos^2 \theta\right)\right)\sin\left(\frac{k^2}{m a \mathcal{H}}\right)\right].\;\;\;\;\;\;
\end{eqnarray}
Thus for the $i=a$ components we can write for the gravitational slip:
\begin{eqnarray}
\frac{\Psi( \eta, \vec{k} ) - \Phi( \eta, \vec{k} )}{\Phi( \eta, \vec{k} )} =\frac{k^2}{2m^2a^2} (1+\cos^2\theta)
\end{eqnarray}
which, as commented before, is $\Od(\epsilon)$ in the adiabatic expansion.

As expected, the previous expressions smoothly tend to the standard 
CDM behaviour discussed in the previous section for $k^2/(ma{\cal H})\ll 1$ (see Fig. \ref{FigScalar}), indeed
\begin{eqnarray}
\delta A_{c a}( \eta, \vec{k} ) &\simeq& -3 C_{a 2}( \vec{k} ) a^{-1/2} \frac{m^2 a^2 \mathcal{H}^2}{k^4} \propto \eta\;;
\\
\Phi( \eta, \vec{k} ) &\simeq& \Psi( \eta, \vec{k} ) \simeq \sqrt{\frac{8\pi G}{3}}\frac{9}{\sqrt{2}}\; C_{a 2}( \vec{k} )  a^{-1/2} \frac{m^3 a^2 \mathcal{H}^3}{k^6} \propto \text{constant}\;,
\\
\delta ( \eta, \vec{k} ) =\frac{\delta \rho( \eta, \vec{k} )}{\rho} &\simeq& \sqrt{\frac{8\pi G}{3}}\frac{m}{a^{1/2}\mathcal{H}} \sqrt{2}     \left( - \frac{3 m^2 a^2 \mathcal{H}^2}{k^4}\right)C_{a 2}( \vec{k} )\propto a\;,
\end{eqnarray}

\begin{figure}
\includegraphics[width = 0.65 \textwidth]{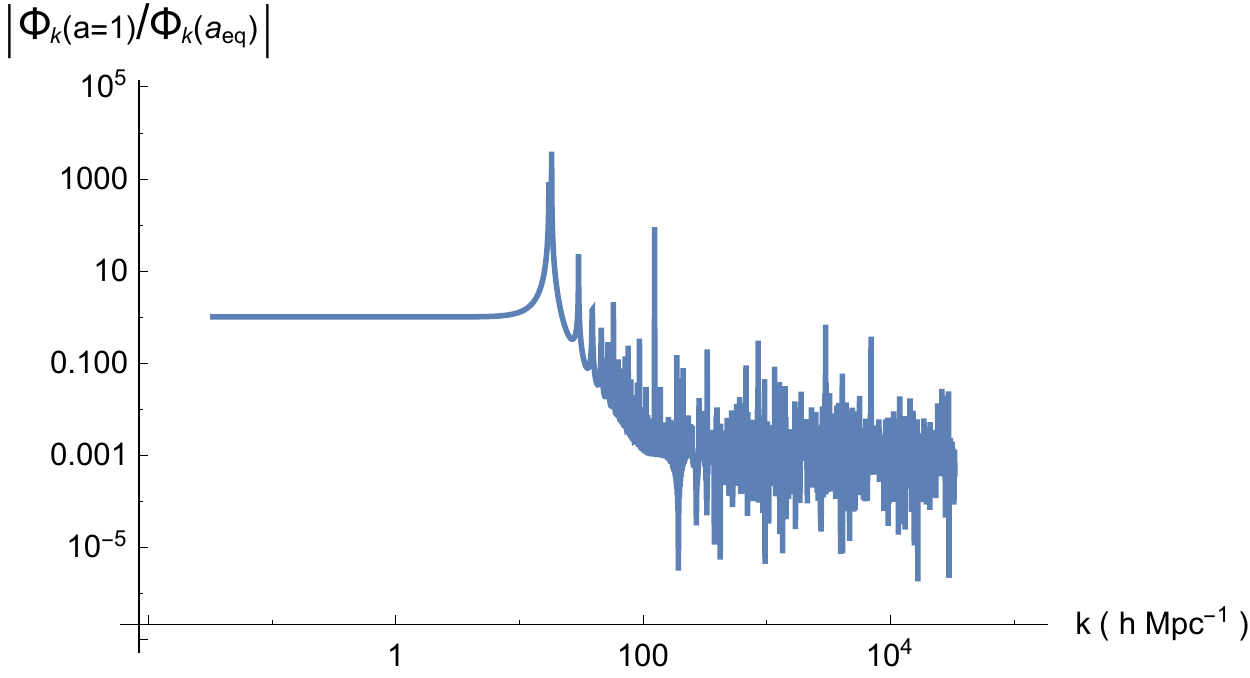}
\caption{Expected correction in the linear transfer function for a vector mass  
 $m=10^{-22}$ eV. We see that the suppression is relevant for 
 $k>10 h$ Mpc$^{-1}$ }\label{transfer}
\end{figure}

In the opposite limit $k^2/(ma{\cal H})\gg 1$ 
expressions (\ref{solk})-(\ref{deltapvec}) imply that the perturbations
\begin{eqnarray}
\delta A_{c a}( \eta, \vec{k} ) &\simeq& a^{-1/2} \left( C_{a 2}( \vec{k} )\; \cos \left(\frac{k^2}{m a \mathcal{H}}\right) + C_{a 1}( \vec{k} )\; \sin \left(\frac{k^2}{m a \mathcal{H}}\right) \right)\;,
\\
\Phi( \eta, \vec{k} ) &\simeq& \Psi( \eta, \vec{k} ) \simeq - \sqrt{\frac{8\pi G}{3}}\frac{3}{\sqrt{2}} \; \frac{m \mathcal{H}}{k^2a^{1/2}}\; \left( C_{a 2}( \vec{k} )\; \cos \left(\frac{k^2}{m a \mathcal{H}}\right) + C_{a 1}( \vec{k} )\; \sin\left(\frac{k^2}{m a \mathcal{H}}\right) \right)\;,
\\
\delta\rho( \eta, \vec{k} ) &\simeq& \sqrt{\frac{3}{8\pi G}}\sqrt{2}\;\frac{m \mathcal{H}}{a^{5/2}}\;\;\left( C_{a 2}( \vec{k} )\; \cos\left(\frac{k^2}{m a \mathcal{H}}\right)+ C_{a 1}( \vec{k} )\; \sin\left(\frac{k^2}{m a \mathcal{H}}\right)\right)\;,
\end{eqnarray}
are all oscillating and decaying (see Fig. \ref{FigScalar}). This is completely analogous to the 
scalar field DM case, and this is the reason why on scales 
with $k^2\gg ma{\cal H}$ we expect a suppression in the matter power spectrum as compared to the standard CDM. See Fig. \ref{transfer}
for the modification on the linear transfer function $\Phi_k(a_0)/\Phi_k(a_{eq})$
induced on small scales. Notice however 
that the possibility of generating a  gravitational 
slip is absent in the scalar field case.

\section{Vector perturbations: results}

From equation (\ref{Eq0iVec}) ,
\begin{eqnarray}
\vec{Q}( \eta, \vec{k} ) &=& \left\lfloor- \frac{6}{k^2 a^2} i \left(\vec{k}\dot{\vec{A}} \left(\hat{k}\left(\vec{k}\vec{\delta A}\right) -\vec{\delta A}\right) + m^2 a^2 \left(\vec{Q}\vec{A} - \delta A_0\right) \left(\hat{k}\left(\vec{k}\vec{A}\right) - \vec{A}\right)\right)\right\rfloor\;, \label{Q}
\end{eqnarray}
we get to the leading order, the following results in the two regimes in 
which we are interested

\subsection{Particle regime ($k\sim {\cal H}$, i.e.  $k/(ma) \sim \epsilon$)}

We see that to the leading order, only the $i=p,\; pk$ components contribute to the vector modes. Such modes did not contribute to the scalar perturbations in this regime, and accordingly they are purely vector like. Thus we get
\begin{eqnarray}
\vec{Q}( \eta, \vec{k} ) &=& \sqrt{\frac{8\pi G}{3}} \frac{3i \sqrt{2} \mathcal{H}}{k a^{3/2}}\left[ \sin(\theta)  C_{p,s}(  \vec{k} )\;\hat{u}_a -\cos(\theta)C_{p,s}(  \vec{k} ) \;\hat{u}_p +\cos(\theta) C_{pk,s}(  \vec{k} )\; \hat{u}_{pk}\right] \; .
\end{eqnarray}
All the components decay as $a^{-2}$ in the matter dominated era. This is the same behaviour expected for vector modes in standard CDM.
Notice also that only the sine components of the vector perturbations
$\delta \vec A_s$ actually contribute to the vector modes. This can be understood since being the background $\vec A$ a cosine function,
the first term in (\ref{Q}), which is the only one contributing to the leading order, contains a $\dot{\vec A}$ factor which  
 in the average procedure is only non-vanishing for sine perturbations.

\subsection{Wave regime ($k\sim ({\cal H}ma)^{1/2}$ i.e.  $k/(ma)\sim \epsilon^{1/2}$)}
In this regime also only the $i=p,\; pk$ components contribute to the vector modes
\begin{eqnarray}
\vec Q( \eta, \vec{k} ) &=&\sqrt{\frac{8\pi G}{3}} \frac{3i \sqrt{2} \mathcal{H}}{k a^{3/2}}\left[-\sin(\theta)\left( C_{p 1}( \vec{k} ) \cos\left(\frac{k^2}{ ma {\cal H}}\right) - C_{p 2}(  \vec{k} ) \sin\left(\frac{k^2}{ ma {\cal H}}\right)\right)\; \hat{u}_a\right.\nonumber \\
&+&\left.\cos(\theta)\left( C_{p 1}( \vec{k} ) \cos\left(\frac{k^2}{ ma {\cal H}}\right) - C_{p 2}(  \vec{k} ) \sin\left(\frac{k^2}{ ma {\cal H}}\right)\right) \; \hat{u}_p\right.
\nonumber \\
&+&\left. \cos(\theta)\left( C_{pk 1}( \vec{k} ) \cos\left(\frac{k^2}{ ma {\cal H}}\right) - C_{pk 2}( \vec{k} ) \sin\left(\frac{k^2}{ ma {\cal H}}\right)\right)\;\hat{u}_{pk} \right].
\end{eqnarray}
We see that once again all the components decay as $a^{-2}$
but with an oscillating behaviour. Also in this case  only the $\delta \vec A_s$ actually contribute in the average.

In both regimes, even though we have a source for the 
vector modes, they actually decay in the same fashion as in standard CDM, 
so that we do not expect large contributions at late times, unless they were produced with very large initial amplitudes.

To summarize this section, we have seen that in both regimes the $i=a$ component contributes to the scalar but not to the vector perturbations, whereas for the $i=p, \; pk$ components the situation is the other 
way around, contributing to the vector perturbations only. In addition, 
 in the $k\sim {\cal H}$ regime perturbations behave exactly as 
in standard CDM, whereas in the $k\sim ({\cal H}ma)^{1/2}$ regime
we find a different behaviour implying that all the scalar perturbations
decay with expansion in the same way as in scalar field DM, 
but unlike the scalar case, a small but non-vanishing gravitational slip 
is generated for the $i=a$ component.

\section{Tensor perturbations}
So far we have neglected the tensor perturbations in the average equations.
In order to extract the equations for such modes we use the projector
\begin{eqnarray}
\Lambda_{ij, lm} \equiv \left( P_{i l}P_{j m}-\frac{1}{2} P_{i j} P_{l m} \right)\;;\;\; P_{i j} \equiv \delta_{i j} - \hat{k}_i \hat{k}_j\;. 
\end{eqnarray}
Thus, contracting with Einstein equations we obtain
\begin{eqnarray}
\Lambda_{ij, l m} E^l_m \equiv \Lambda_{ij, lm} \left( \delta G^l_m - 8 \pi G \left\lfloor\delta T^l_m\right\rfloor \right) &=& E^i_j - \hat{k}_i\hat{k}_l E^l_j - \hat{k}_j \hat{k}_m E^i_m
\\
&+& \hat{k}_i\hat{k}_j \left(\hat{k}_l\hat{k}_m E^l_m\right) - \frac{1}{2} \left(\delta_{ij} - \hat{k}_i \hat{k}_j\right)\left(Tr\left(E^l_m\right) - \hat{k}_l\hat{k}_m E^l_m \right)\;.\nonumber
 \end{eqnarray}

We will calculate the components of the tensor perturbation in the  orthonormal basis defined by: $\left\{\hat{u}_{1}= \hat{u}_{pk}, \hat{u}_{2}=\cos \theta\; \hat u_p-\sin\theta\; \hat u_a \; , \hat{u}_{3}=\hat k=\sin \theta\; \hat u_p+\cos\theta\; \hat u_a \right\}$. In this basis the 
tensor perturbation takes the standard form,
\begin{eqnarray}
h_{i j}( \eta, \vec{k} ) \equiv \left(\begin{array}{ccc}
h_+ & h_{\times}& 0\\
h_{\times}& -h_{+}&0\\
0&0& 0
\end{array}\right)\;.
\end{eqnarray}

From the projection of Einstein equations we reach
\begin{eqnarray}
\ddot{h}_{(+,\times)} &+& 2 \mathcal{H}\dot{h}_{(+,\times)} + \left(k^2 - 2 \left(\mathcal{H}^2 + 2 \dot{\mathcal{H}}\right) \sin^2(\theta) \right) h_{\times} =  S_{(+,\times)}\;,\nonumber
\end{eqnarray}
where
\begin{eqnarray}
S_+( \eta, \vec{k} )&=& -16 \pi G \sin(\theta)\left\lfloor\sin(\theta) \left( - \frac{\Psi}{a^2} \left(\dot{A}^2 - m^2 a^2 A^2\right) 
+\frac{\Phi}{a^2} \dot{A}^2 - \frac{\dot{A}\dot{\delta A}_a}{a^2} - i \frac{k \dot{A}}{a^2} \delta A_0 \cos(\theta) + m^2  A \delta A_a\right)\right.\nonumber
\\
&-& \left. \cos(\theta) \left( \frac{\dot{\delta A}_p \dot{A}}{a^2} + i \frac{k}{a^2} \sin(\theta) \dot{A} \dot{\delta A}_0 - m^2 \delta A_p A\right)\right\rfloor,\\
S_{\times}( \eta, \vec{k} )&=&16 \pi G \sin(\theta) \zerof{\frac{\dot{\delta A}_{pk} \dot{A}}{a^2} - m^2 A \delta A_{pk}}.
\end{eqnarray}

\subsection{Particle regime ($k\sim {\cal H}$, i.e.  $k/(ma) \sim \epsilon$)}

In this regime the sources vanish to the 
leading order
\begin{eqnarray}
S_{+,\times}( \eta, \vec{k} ) = 0
\end{eqnarray}
so that the generation of gravity waves  will be negligible.
\subsection{Wave regime ($k\sim ({\cal H}ma)^{1/2}$ i.e.  $k/(ma)\sim \epsilon^{1/2}$)}
 In this regime, the average sources read
\begin{eqnarray}
S_{+}( \eta, \vec{k} ) &=& -\sqrt{\frac{8\pi G}{3}}\frac{3\sin(\theta)}{a^{3/2}\sqrt{2}}\frac{k^2 \mathcal{H}}{m a} \left( (4\cos^2( \theta)-1)\sin(\theta)    \left[\left(C_{a 2}( \vec{k} ) \left(1 - \frac{3 m^2 a^2 \mathcal{H}^2}{k^4}\right) + 3 C_{a 1}( \vec{k} ) \frac{m a \mathcal{H}}{k^2}\right) \cos \left(\frac{k^2}{m a \mathcal{H}}\right) \right. \right.
\nonumber \\
 &+&\left.\left. \left(C_{a 1}( \vec{k} ) \left(1 - \frac{3 m^2 a^2 \mathcal{H}^2}{k^4}\right) - 3 C_{a 2}( \vec{k} ) \frac{m a \mathcal{H}}{k^2}\right) \sin \left(\frac{k^2}{m a \mathcal{H}}\right)\right]\right. \nonumber \\
& +&\left.  \left(4 \sin^2(\theta)-1\right) \cos(\theta)  \left[C_{p 1}( \vec{k} ) \sin\left(\frac{k^2}{m a \mathcal{H}}\right)+ C_{p 2}( \vec{k} ) \cos\left(\frac{k^2}{m a \mathcal{H}}\right)\right]\right)\;, \label{S+}
\\
S_{\times}( \eta, \vec{k} ) &=& \sqrt{\frac{8\pi G}{3}}\frac{3\sin(\theta)}{ a^{3/2}\sqrt{2}} \frac{k^2 \mathcal{H}}{m a}  \left(C_{pk 1}( \vec{k} ) \sin\left(\frac{k^2}{m a \mathcal{H}}\right)+ C_{pk 2}( \vec{k} ) \cos\left(\frac{k^2}{m a \mathcal{H}}\right)\right)\;.\label{Sx}
\end{eqnarray}
Thus we see that the $i=a$ modes associated to the scalar 
perturbations are not purely scalar, but rather scalar-tensor modes whose tensor components have
only $+$ polarization. The vector perturbations with $i=p$ are actually vector-tensor modes 
also with $+$ polarization  and finally the $i=pk$
component generates vector-tensor perturbations with $\times$ polarization.

Redefining the field as $h_{(+,\times)}=a^{-1}\tilde{h}_{(+,\times)}$, we can obtain solutions in terms of the Green's functions:
\begin{eqnarray}
\ddot{\tilde{h}}_{(+,\times)} + k^2\tilde{h}_{(+,\times)} = a S_{(+,\times)}\;,
\end{eqnarray}
with solution,
\begin{eqnarray}
h_{(+,\times)}( \eta, \vec{k} ) = \frac{1}{a} \int^\eta_{\eta_1} G\left(\eta - \eta'\right) a(\eta') S_{(+,\times)}(\eta',\vec{k}) d\eta'= \frac{1}{a} \int^\eta_{\eta_1} \frac{\sin\left(k \left(\eta - \eta'\right)\right)}{2 k} a(\eta') S_{(+,\times)}(\eta', \vec{k}) d\eta' \;. \label{solGW}
\end{eqnarray}

Let us consider for example $h_{\times}$ and assume $C_{pk 1} =0$
\begin{eqnarray}
h_{\times}( \eta, \vec{k} ) =\sqrt{\frac{8\pi G}{3}} \frac{3 \sin(\theta) }{2\sqrt{2} } \frac{k }{ma(\eta) } C_{pk 2}( \vec{k} ) \int^\eta_{\eta_1} d\eta' \sin\left(k\left(\eta-\eta'\right)\right) \left(\frac{{\cal H}(\eta')}{a^{3/2}(\eta')}\right) \cos\left(\frac{k^2}{ ma(\eta'){\cal H}(\eta')} \right)\;.
\end{eqnarray}

In this regime $k\gg {\cal H}$ so that $\sin(k(\eta-\eta'))$ in the integrand
oscillates rapidly whereas $S_{(+,\times)}(\eta')$ evolves slowly with time, so that can use partial integration as in (\ref{pInt}) so that the leading term will be 
\begin{eqnarray}
h_{\times}( \eta, \vec{k} ) = \sqrt{\frac{8\pi G}{3}} \frac{3 \sin(\theta) }{2\sqrt{2} } \frac{C_{pk 2}( \vec{k} )}{ma(\eta) } \left[\cos\left(k\left(\eta-\eta'\right)\right) \left(\frac{{\cal H}(\eta')}{a^{3/2}(\eta')}\right) \cos\left(\frac{k^2}{ ma(\eta'){\cal H}(\eta')} \right)\right]_{\eta_1}^\eta \;.\label{h12}
\end{eqnarray}
Thus, we obtain waves with an amplitude that 
decays as $a^{-1}$ and propagate at the speed of light.
A completely analogous result can be obtained for the $h_+$ polarization.

\section{Gravitational wave detection}
As shown above, scalar perturbations given by the $C_{a,(1,2)}$ components generate a gravity wave background with $h_+$ polarization in the $k\sim ({\cal H}ma)^{1/2}$ regime. If all the 
cosmological DM is generated by the vector field, it is possible 
to estimate the spectrum of gravity waves associated to such components and 
compare with the sensitivity of present and future detectors. 

Note that for the mentioned components in this regime,  the source $S_+$  in (\ref{S+}) can be written in terms of the scalar perturbation $\Phi$ given in (\ref{Phik})
as
\begin{eqnarray}
S_+(\eta, \vec{k})=\sin^2(\theta)\left(4 \cos^2(\theta) -1\right) \frac{k^4}{m^2 a^2} \Phi(\eta, \vec{k}).
\end{eqnarray}  
Thus, from (\ref{solGW}), integrating by parts as in the previous section we get:
\begin{eqnarray}
h_{+}(\eta, \vec{k}) = \frac{k^2}{2m^2a(\eta)}\sin^2(\theta)\left(4 \cos^2(\theta) -1\right)  \left[ \cos\left(k \left(\eta - \eta'\right)\right) a^{-1}(\eta') \Phi(\eta',\vec{k})\right]_{\eta_{eq}}^\eta
\end{eqnarray}
where in order to simplify the calculation we have assumed an instantaneous change to matter domination at equality, $\Omega_m(a_{\text{eq}}) = \Omega_{\text{rad}}(a_{\text{eq}})$, for both background and perturbations.
Thus, we set the initial amplitude of the gravity waves to zero at 
equality. The term in brackets oscillates with an amplitude that 
decays as $a^{-2}$ so that it is dominated by the lower integration time. Evaluating it at $\eta'=\eta_{eq}$, we obtain
\begin{eqnarray}
h_{+}(\eta,\vec{k}) = \frac{k^2}{2m^2a(\eta)}\sin^2(\theta)\left(1-4 \cos^2(\theta) \right)  \cos\left(k \left(\eta - \eta_{eq}\right)\right) a_{eq}^{-1} \Phi(\eta_{eq},\vec{k}).
\end{eqnarray}
We can see at equality when the amplitude of the 
gravity wave is largest, we have  
\begin{eqnarray}
\frac{h_{+}(\eta_{eq},\vec{k})}{\Phi(\eta_{eq},\vec{k})} = \frac{k^2}{2m^2a_{eq}^2}\sin^2(\theta)\left(1-4 \cos^2(\theta) \right)
\end{eqnarray}
which is $\Od(\epsilon)$. This is the reason why we could neglect the
contribution of tensor modes in the evolution of scalar and vector perturbations in Section V.  

In this regime, with $k\gg k_{eq}$ where $k_{eq}={\cal H}_{eq}=0.073\, \Omega_mh^2$ Mpc$^{-1}$
it is possible to obtain $\Phi(\eta_{eq},\vec{k})$ directly  from the linear scalar transfer function \cite{Dodelson}
\begin{eqnarray}
\Phi(\eta_{\text{eq}},\vec{k}) = \frac{9}{10} \Phi^{\text{{\tiny  prim}}} T(k) \simeq \frac{9}{10}\Phi^{\text{{\tiny  prim}}} \frac{12 k_{\text{eq}}^2}{k^2} \; \ln\left(\frac{k}{8k_{\text{eq}}}\right),
\end{eqnarray}
with $\Phi^{\text{{\tiny  prim}}}$  the primordial amplitude of 
perturbations generated during inflation, with a 
spectrum
\begin{eqnarray}
P_\Phi(k)=\frac{k^3}{2\pi^2}\vert \Phi^{\text{{\tiny  prim}}}\vert^2=A_s\left(\frac{k}{k_0}\right)^{n_s-1}
\end{eqnarray}
where the values  of the parameters, $\ln(10^{10} A_s)=3.089 \pm 0.036$, $n_s=0.9655  \pm 0.0062$ and the pivot scale $k_0=0.05$ Mpc$^{-1}$  correspond to Planck observations \cite{Planckinf}. 

Finally, the energy density today of gravitational waves with $+$ polarization per energy interval and solid angle unit reads \cite{defOmGW,defOmGWb},
\begin{eqnarray}
\frac{d\Omega_{\text{GW}} (k,\eta_0)}{d\Omega} = \frac{1}{\rho_c}\frac{d^2 \rho_{\text{GW}}}{d\Omega d \ln(k)} = \frac{k^3 \vert\dot{h_+}\vert^2}{48 \pi^3 H_0^2}.
\end{eqnarray}

Integrating over the whole solid angle we obtain the 
spectral energy density,
\begin{eqnarray}
\Omega_{\text{GW}} (k, \eta_0) &=& \int d\Omega \frac{k^5 \vert h_+\vert^2}{48 \pi^3 H_0^2} = 1.605A_s \frac{k^2}{H_0^2} \left(\frac{k_{\text{eq}}^2}{m^2 a_{\text{eq}}} \ln\left(\frac{k}{8k_{\text{eq}}}\right)\right)^2 \left(\frac{k}{k_0}\right)^{n_s-1}\;, \;\; k_{eq}\ll k\ll ma_{eq}.
\end{eqnarray}

\begin{figure}[t]
\includegraphics[width=0.75\textwidth]{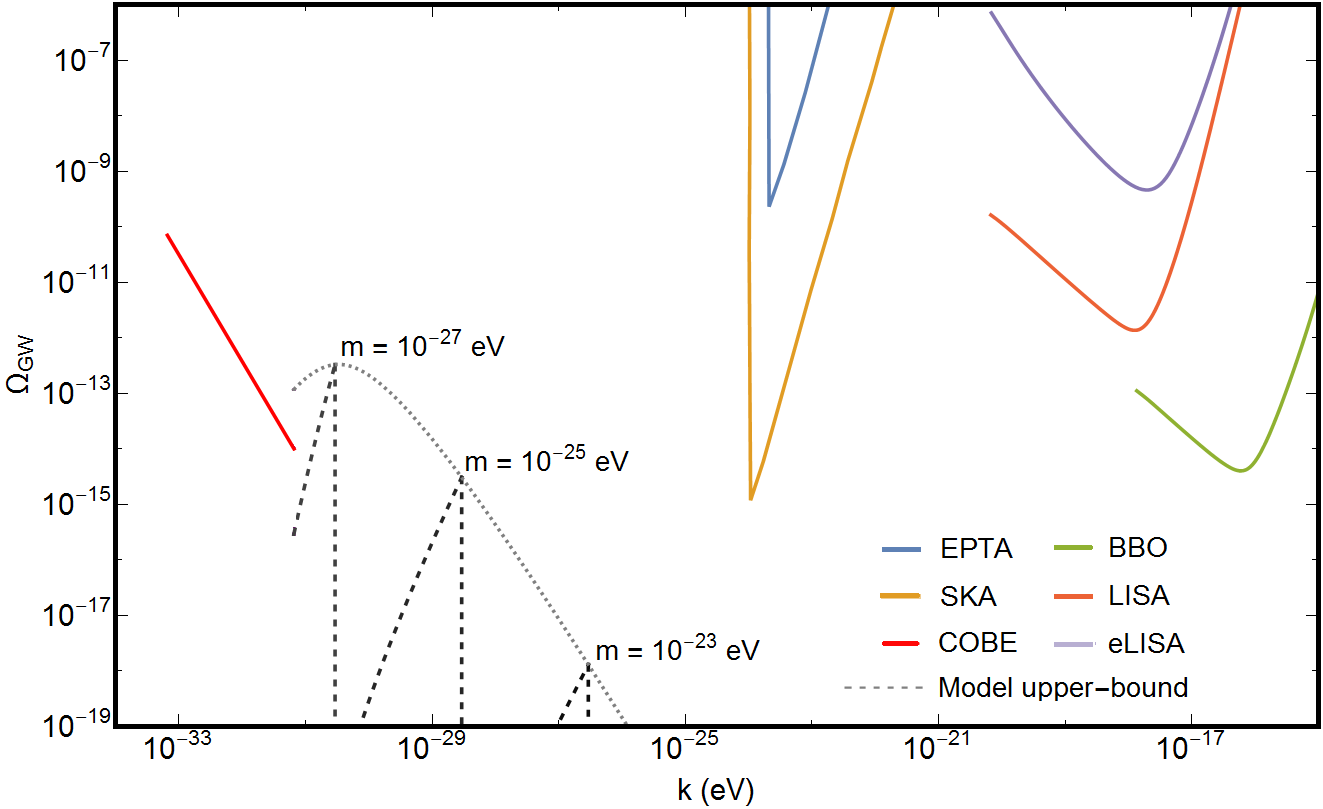}
\caption{In this figure the sensitivities to the energy density abundance of gravitational waves per mode today of COBE (red), EPTA (blue), SKA (orange), BBO (green), LISA (orange) and eLISA (purple)  are plotted \cite{plotGW,plotGWb}. The black dashed line show the upper bound limit of the massive vector gravitational waves production, as it can be seen its detection is unlikely with the future detectors.}\label{GW}
\end{figure}

In Fig. \ref{GW}, we compare the prediction of the vector field DM model with  the sensitivity of present and future gravity wave experiments. The best sensitivity  at low frequencies correspond to the CMB data, but unfortunately the spectral range only reaches $k=k_{eq}$ just in the
limit of the wave regime. On the other hand, at higher frequencies,    
SKA pulsar timing limit is twelve orders of magnitude above the production 
prediction.

\begin{figure}[t]
\includegraphics[width=0.65\textwidth]{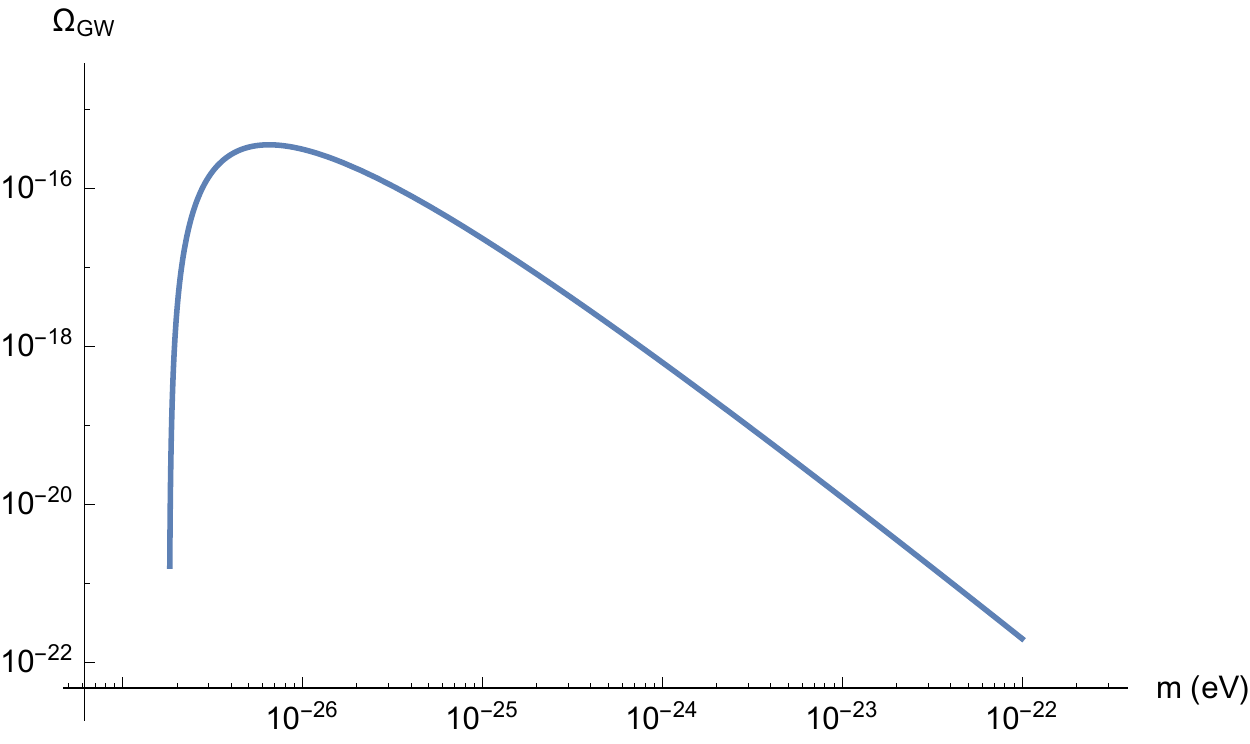}
\caption{Gravity wave abundance as a function of the vector mass. The maximum is expected for $m\simeq 10^{-27}$ eV}\label{OGW}
\end{figure}

If we integrate over $k$ in the gravitational wave production band we get
\begin{eqnarray}
\Omega_{\text{GW}}(\eta_0) =\int_{8k_{eq}}^{ma_{eq}} \frac{dk}{k} \Omega_{GW}(k,\eta_0)=
1.605A_s \frac{k_{eq}^4}{H_0^2 m^4 a_{eq}^2}\left[\frac{k^2}{4}-\frac{k^2}{2}\ln\left(\frac{k}{8k_{eq}}\right)+\frac{k^2}{2}\ln^2\left(\frac{k}{8k_{eq}}\right) \right]_{8k_{eq}}^{ma_{eq}}.
\end{eqnarray}
where we have approximated $n_s=1$ for simplicity. In Fig. \ref{OGW}
we plot $\Omega_{GW}(\eta_0)$ as a function of $m$. 
The expected sensitivity for the combined analysis of the future COrE and Euclid missions \cite{Ngw} on the effective number of relativistic degrees of freedom can be translated into a limit on gravity wave abundance
\begin{eqnarray}
\Omega_{\text{GW}}^{\text{{\footnotesize COrE+Euclid}}}(\eta_0) < 7.6 \;\times 10^{- 8}.
\end{eqnarray}
As can be seen in Fig.\ref{OGW} the maximum production corresponds 
to masses $m \sim 10^{-27}$ eV, but still it is a few orders of magnitude below
the mentioned limit.

\section{Conclusions}
Ultralight bosonic fields are natural DM candidates which
can avoid some of the small-scale problems of the standard CDM
model. Most of the work developed so far in this field has focused on
the simplest implementation of this scenario based on  
scalar fields. In this work we have considered  the case of
ultralight vector fields. 

The first difficulty in the higher spin case already appears at 
the background level, since such fields generically break isotropy. Fortunately, a general result \cite{GIsTh} shows that for massive 
fields without self interactions and masses much larger than the expansion rate, the average energy-momentum tensor 
is isotropic and behaves as CDM. 

\begin{figure*}[h]
\includegraphics[width=\textwidth]{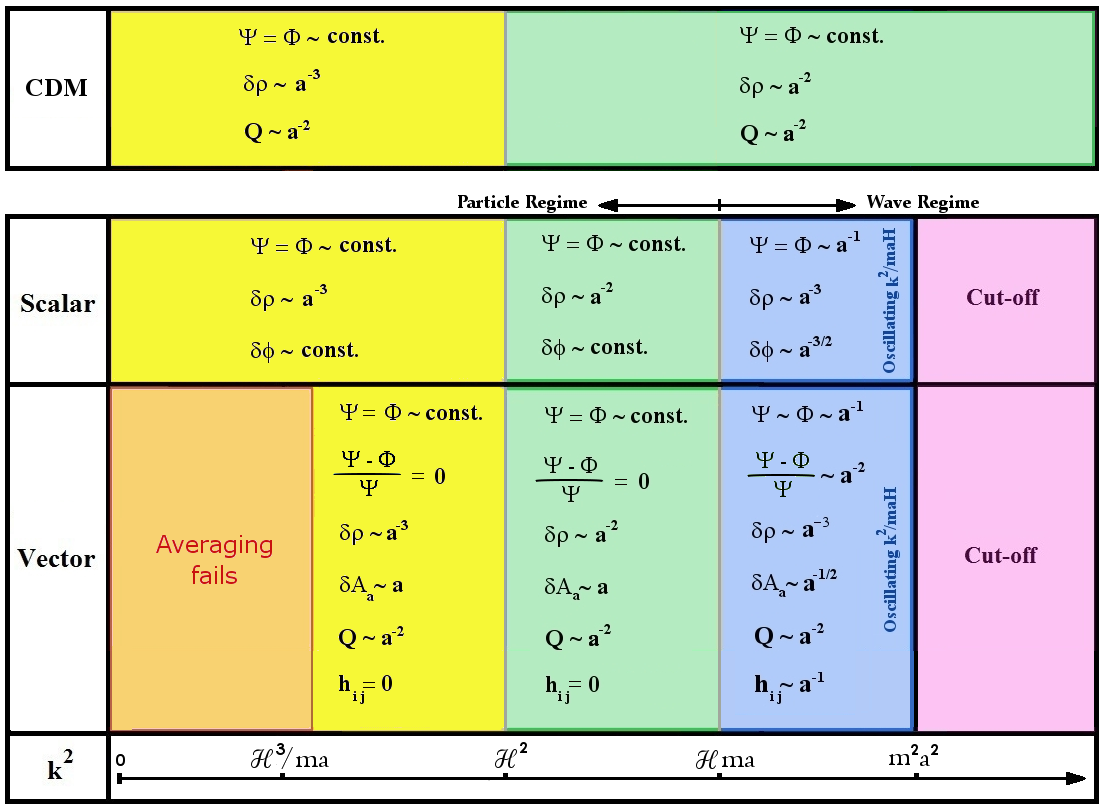}
\caption{The diagram shows the leading order behaviour of perturbations sourced by the different types of DM. In the standard CDM scenario, DM can be the source of scalar and vector perturbations. In this case there are only two relevant regimes, namely, sub- and super-Hubble modes. 
Coherent scalar DM can only source scalar perturbations, but their scaling 
depends not only on the wavenumber but also on the mass of the field, 
defining a total of four different regimes. For coherent vectors, we have the same regimes as for the scalar, but in this case vector
and tensor perturbations can also be sourced, as well as a gravitational slip. Both massive scalar and vector fields mimic CDM in yellow and green regions.}\label{Summary}
\end{figure*}

At the perturbation level, we have considered an adiabatic expansion in two different regimes, the so called particle regime with
 $k/(ma) \sim \epsilon$ and the wave regime with $k/(ma)\sim \epsilon^{1/2}$. Very much as in the scalar case, we find that for vectors, the particle regime is indistinguishable from CDM. However, in the wave regime  important differences with respect to the scalar case arise. Thus,
 perturbations in the vector field support three kinds of metric perturbations. On one hand, we have two scalar modes with a small but non-vanishing  sound speed $c_s^2\sim k^2/(m^2a^2)$ which suppresses structure formation for 
 $k>\sqrt{{\cal H}ma}$. Such modes generate a non-vanishing gravitation
 slip of order $(\Phi-\Psi)/\Phi\sim k^2/(m^2a^2)$ which is a specific feature of this ultralight vector field DM model.  
 In addition, the scalar modes source tensor modes 
 with a small amplitude $h/\Phi\sim k^2/(m^2a^2)$ and a characteristic 
 spectrum which peaks around $k_{max}\sim ma_{eq}$. The amplitude of this
 gravity wave spectrum is however below the sensitivity of present and future detectors. Nevertheless, the calculation done in this work has been conservative in the sense that  we have focused only in the 
 potential generation of gravity waves in the matter dominated era.
 A complete study would require to consider also  perturbations in the radiation era. On the other hand, we have four vector modes which are also 
 sources of gravity waves. The vector modes decay as  $a^{-2}$, i.e. in 
 a similar fashion as in standard CDM cosmologies, so that we expect a negligible amount of vector modes at late times also in this model. 
  In Fig.\ref{Summary} a summary of the perturbations behaviour in the different regions of the spectrum is shown for massive vector and scalar models.

\vspace{0.2cm}
{\bf Acknowledgements:}
 This work has been supported by the MINECO (Spain) projects FIS2014-52837-P, FPA2014-53375-C2-1-P,
 and Consolider-Ingenio MULTIDARK CSD2009-00064.

\end{document}